\documentclass[twocolumn]{aa}
\usepackage{graphicx,natbib}

\def\Epk{E_{\rm p}}

\def\Dt{\Delta t}
\def\F0{F_{\rm 0}}
\def\E00{E_{\rm 0}}

\def\t0{t_{\rm 0}}

\def\Liso{L/\Omega}
\def\Tj{\theta_{\rm j}}
\def\e1{\eta-1}
\def\P0{\Phi_{\rm 0}}

\def\lag{\Delta t}

\def\tr{t_{\rm r}}

\def\S{Sect. }

\newcommand{\ltsima} {\; \buildrel < \over \sim \;}
\newcommand{\gtsima} {$\; \buildrel > \over \sim \;$}
\newcommand{\lta} {\lower.5ex\hbox{\ltsima}}
\newcommand{\gta} {\lower.5ex\hbox{\gtsima}}

\begin{document}

\title{Interpretations of gamma-ray burst spectroscopy\\ II. Bright BATSE bursts}

\author{F. Ryde \inst{1}
 \and D. Kocevski \inst{2}
\and Z. Bagoly \inst{3} \and  N. Ryde \inst{4} \and A.
M\'esz\'aros \inst{1,5}}

 \offprints{F. Ryde}
 \institute{Stockholm Observatory, AlbaNova
University Center, SE-106 91 Stockholm, Sweden
 \and Department of Physics and Astronomy, Rice University, Houston, TX 77005, USA
\and
 Laboratory for Information Technology, E\"otv\"os
University, Budapest, 1117, Hungary
 \and Uppsala Astronomical Observatory, Box 515, 751 20 Uppsala, Sweden
\and Astronomical Institute of the Charles University, 180 00
Prague 8, V Hole\v{s}ovi\v{c}k\'ach 2, Czech Republic}
  \date{Received 30 June 2004 / Accepted 8 November 2004}

 \abstract{
We analyze the spectral lags of a sample of bright gamma-ray burst
{\it pulses} observed by {\it CGRO} BATSE and compare these with
the results of high-resolution spectroscopical investigations. We
find that pulses with hard spectra have the largest lags, and that
there is a similar, but weaker correlation between
hardness-intensity correlation index, $\eta$, and lag. We also
find that the lags differ considerably between pulses within a
burst. Furthermore, the peak energy mainly decreases with
increasing lag. Assuming a lag-luminosity relation as suggested by
Norris et al., there will thus be a positive
luminosity--peak-energy correlation. We also find that the
hardness ratio, of the total flux in two channels, only weakly
correlates with the spectral evolution parameters. These results
are consistent with those found in the analytical and numerical
analysis in Paper I. Finally, we find that for these bursts,
dominated by a single pulse, there is a correlation between the
observed energy-flux, $F$, and the inverse of the lag, $\lag$: $F
\propto \lag^{-1}$. We interpret this flux-lag relation found as a
consequence of the lag-luminosity relation and that these bursts
have to be relatively narrowly distributed in $z$. However, they
still have to, mainly, lie beyond  $z  \sim 0.01$, since they do
not coincide with the local super-cluster of galaxies. We discuss
the observed correlations within the collapsar model, in which the
collimation of the outflow varies. Both the thermal photospheric
emission as well as non-thermal, optically-thin synchrotron
emission should be important.

   }

\authorrunning{Ryde et al.}
\titlerunning{GRB Spectroscopy}

   \maketitle

\section{Introduction}

 In a previous paper (\citet{ryde05};
Paper I) we demonstrated the relation between the two approaches
used to characterize the spectral evolution in gamma-ray bursts
(GRBs): the one using spectral lags between energy channels and
the one using high-resolution data, leading to detailed
parametrization of the evolution. Specifically, we investigated
this relation for individual pulses, using an analytical model and
numerical simulations. We  showed that the spectral evolution
described by high spectral-resolution data, in empirical
correlations, naturally leads to correlations involving spectral
lags. The relation between the analyses is not trivial and a
general description can only be given approximately, but can be
given exactly case by case. This is because the lag measures a
combination of spectral evolution parameters. It further depends
on the relative channel widths the data are divided into.

We will here briefly summarize the main results of the analytical
and numerical simulation analysis from Paper I.  We found that the
spectral lag correlates strongest with the decay time-scale of the
pulse (or equivalently of the peak-energy decay). Bursts with hard
spectra, that is, with large low-energy spectral power-law slope,
$\alpha$, exhibit the largest lags. Similarly, a more energetic
burst, as measured by the peak energy of the spectra, also, most
often, yields a large lag. However, for very soft bursts the
relation is the opposite. Furthermore, the ratio of the flux in
two different energy channels, the hardness ratio, is found to
only weakly vary with $\alpha$ and $\eta$ (coefficient of the
hardness-intensity correlation, HIC, see Eq. 1 in Paper I), and it
is mainly determined by how energetic the burst is. Characteristic
differences in the behavior of hard and soft pulses are
identified, as well as differences in behavior between more and
less energetic bursts. Further, we could also reproduce the
observation by \citet{KL03}, namely a good correlation between lag
and the quantity $\Phi_0$ normalized by the peak flux, $= \P0/F$.
The parameter $\Phi_0$ describes the decay rate of the peak energy
of the spectrum (see eq. 2, in Paper I). For different pulses
within a burst, the spectral parameters are, in general, not the
same, and thus the lag is not expected to be the same during the
whole bursts.

In this paper we will analyze GRB pulses, observed by the Burst
and Transient Source Experiment (BATSE) on the {\it Compton
Gamma-Ray Observatory (CGRO)}, and compare the results with the
trends and correlations summarized above. This will be done in \S
\ref{sec:observations}. In particular, we will study the spectral
lags and hardness ratios. In \S \ref{sec:physics}, we will also
discuss how our results relate to the physical models that have
been proposed to explain the observed lag correlations. We
conclude our discussion in \S \ref{sec:discussion}

\section{{\it CGRO} BATSE Observations}
\label{sec:observations}

We defined a sample of strong and well-separated pulses by using
the \cite{BR01} sample which consists of 47 pulses that all show a
clear spectral evolution defined by the hardness-intensity
correlation (HIC) during the pulse decay. From these we chose the
cases which have a 'clean' rise phase that is not contaminated
greatly by previous pulses. We also limited our sample to cases
for which the rise time, $t_r$, is much shorter than the decay
time, $t_d$: $t_r /t_d < 0.2$. This gave us the sample of 17
pulses in 15 bursts, which are presented in Table
\ref{tab:sample}. We analyzed the four-channel discriminator rates
of BATSE and as customary, used channel 1 ($\sim 25-50$ keV) and
channel 3 ($\sim 100-300$ keV). These data are constructed from
three data types from the Large Area Detector (LAD): DISCLA, PREB,
and DISCSC \citep{fish}. We also analyzed the 128-channel data
(high energy-resolution data, HERB), to be able to compare the
results.  We deconvolve the observed count data into photon fluxes
and performed the analysis with RMFIT 1.0b6, provided by the BATSE
team at MSFC (see e.g. \citet{RS00} for details).

\begin{table}

\caption[]{Sample of 17 pulses in 15 GRBs defined by their peak
times}
\begin{flushleft}
\begin{tabular}{lccc}
\\
\hline
\\
{Burst} & {trigger} & {LAD}& {$t_{\rm max}$ [s]}  \\
 \hline
 \vspace{-2mm}
 \\
 GRB911016 &907  &  1    & 1.6           \\
 GRB911031 &973  &  3    & 2.8/24      \\
 GRB911104 &999 & 2    & 4.0       \\
 GRB920525 &1625 &  4    & 4.9        \\
 GRB920830 &1883 &  0    & 1.2             \\
 GRB921207 &2083 &  0    & 8.6      \\
 GRB930201 &2156 &  1    & 14     \\
 GRB941026 &3257 &  0    &  3.0     \\
 GRB950624 &3648 & 3    & 23/41    \\
 GRB950818 &3765 &  1    &  66       \\
 GRB951102 &3891 &  2    & 33      \\
 GRB960530 &5478 &  2    &  2.0     \\
 GRB960804 &5563 &  4 & 1.4             \\
 GRB980125 &6581 &  0    & 48            \\
 GRB990102 &7293 &  6    & 3.2
\\
\hline
\end{tabular}
\end{flushleft}
 \label{tab:sample}
\end{table}

  \begin{figure*}
\centering
\includegraphics[width=\textwidth]{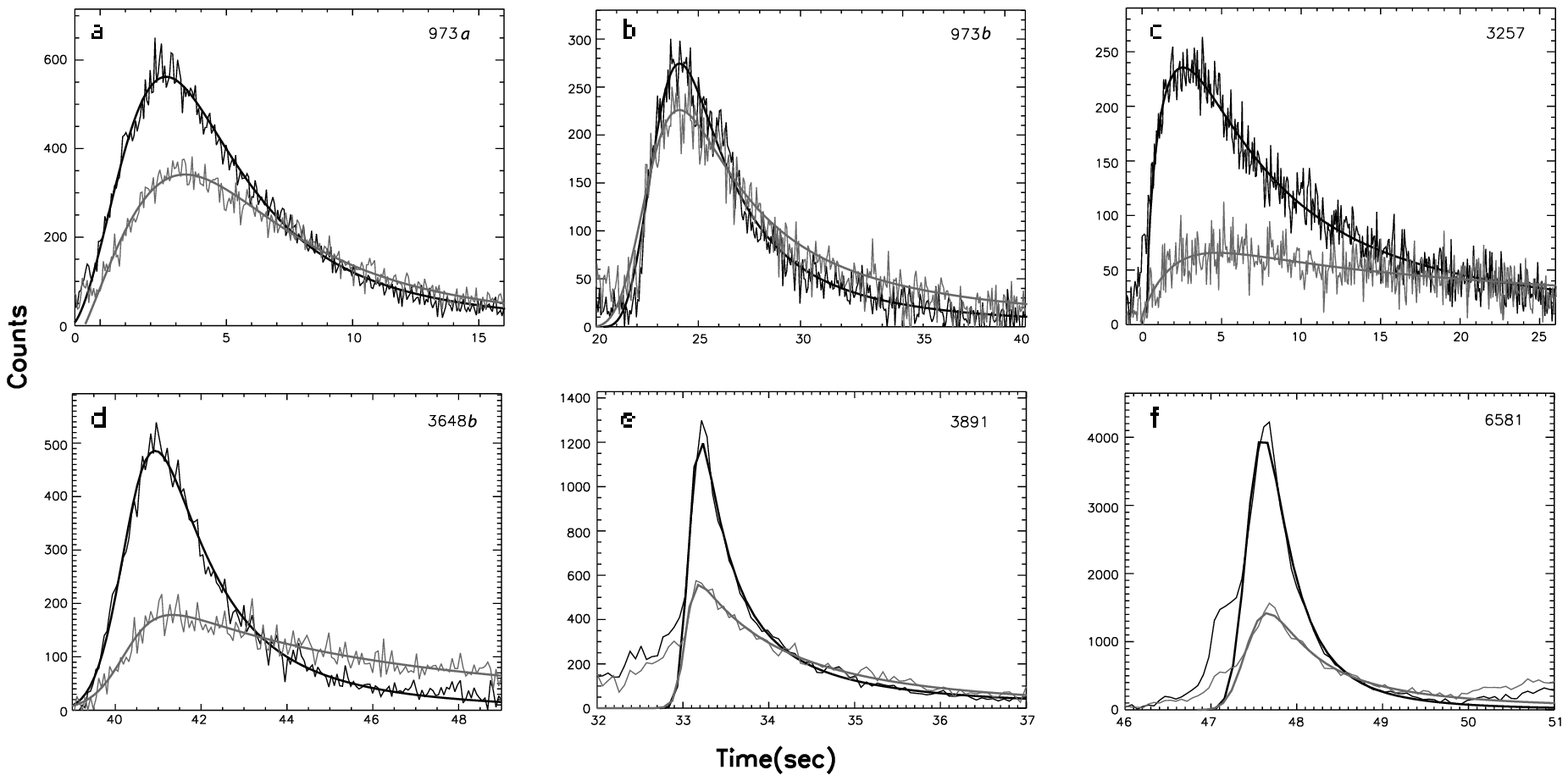}
 \caption{Light-curves of six of the studied pulses;
(a) GRB911031 (\# 973a); (b) GRB911031 (\# 973b); (c) GRB 941026
(\# 3257); (d) GRB960524 (\# 3648);  (e) GRB 951102 (\# 3891); (f)
GRB 980125 (\# 6581). The black curves are for BATSE channel 3 and
the grey curves for channel 1. The smooth line is the best fit
model (Eq. [\ref{F2}]).
 \label{fig:lags}}
 \end{figure*}

\subsection{Analysis of Pulses}

In Paper I, we derived an analytical function for a GRB pulse,
with four parameters (see also \citet{KRL}):
\begin{equation}
F(t)=\frac{A_0 (1-e^{-t/\tr})}{(1+(t+\tr e^{-t/\tr})/\tau)^d}
\label{F2}
\end{equation}
Here $\tau$ is the decay time scale and $\tr$ is connected to the
rise phase. The decay index $d$ is defined by requiring that $F(t)
\rightarrow t^{-d}$ as $t \rightarrow \infty$ and $A_0$ is an
analytical function of the other parameters. We will also use the
peak energy, $\E00$, at the beginning of the pulse, which is given
when $t=0$ in Eq. (\ref{F2}) and defines the onset of pulse
emission according to the analytical model. The channel
light-curves in our sample were fitted with this function. Figure
\ref{fig:lags} depicts a few examples of the studied bursts and
their channels 1 and 3 light curves and the fits are shown by
solid lines. Even though the function was deduced for the
bolometric light curve, it is flexible enough to fit the channel
light-curves, which is reflected by the fact that the fits mainly
have good values of the $\chi^2$. In column 2 of Tab.
\ref{tab:obspeak} the peak times are given for the light curves
over the whole spectral range of BATSE ($\sim 25-2000$ keV). The
peak time is given in seconds since the BATSE trigger, and is thus
different from $t$ in Eq. (\ref{F2}) by a constant offset. The
main purpose of measuring this time is to compare it with the same
time measured on the photon light-curves instead (see \S
\ref{sec:dc})

\subsubsection{Analytic Lags, $lag13$}

For every pulse, the {\it analytic spectral lag}, $\equiv lag13$,
was found by measuring the time difference between the light curve
peaks in channels 1 and 3. These peaks were identified as the
maxima of the function in Eq. (\ref{F2}). These results are given
in the third column in Tab. \ref{tab:obspeak}. We note that there
is a large spread in values of the lag. The largest value is for
the first pulse in GRB960524 (BATSE trigger \# 3648) with 2.65 s.
Furthermore, GRB 980125 (\# 6581) and the second pulse in
GRB911031 (\# 973) are insignificantly different from zero, which
means that the pulses in the four spectral channels peak at the
same time. As noted by \cite{band97} and \cite{norris2}, GRB lags
are, in general, concentrated towards such short time scales ($<$
100 ms)). The case GRB 951102 (\# 3891) is an example of a pulse
with a negative lag, that is, the hardest radiation lags behind
the softer radiation, even though the significance of it being
different from zero is low. Finally, the lags for the two cases in
the sample which have two pulses GRB911031 (\# 973) and GRB960524
(\# 3648) show that the sizes of the lags are significantly
different between the pulses. This is an important observation,
since most often lags are measured over the whole burst and it is
assumed that the lag is constant. Such a lag measurement therefore
more reflects the averaged lag behavior. Furthermore, \cite{BR01}
noted that the HIC index, $\eta$, in multi-pulse bursts are often
constant from pulse to pulse, while \citet{LK96} noted that the
HFC index, $\Phi_0$, is similarly, practically constant. Both
these observations are in contrast to that of the lags, which do
differ significantly. F. Ryde \& D. Kocevski (2004, in prep.) have
made a more detailed analysis of the spectral lag in multi-pulsed
bursts and showed that this is a general feature.

\begin{table*}
\caption[ ]{The peak times  and lags of the pulses in the sample}

\begin{flushleft}
\begin{tabular}{lccccccc}
\hline \noalign{\smallskip}
 &&
\multicolumn{2}{c}{Counts}&& \multicolumn{2}{c}{Photon flux}
\\
\cline{3-4} \cline{6-7} \multicolumn{1}{c}{ Trigger} &&
\multicolumn{1}{c}{peak time [s]} & \multicolumn{1}{c}{$lag13$
[s]} && \multicolumn{1}{c}{peak time [s]} &
\multicolumn{1}{c}{$lag13$ [s]}
  \\
\noalign{\smallskip} \hline \noalign{\smallskip}
907 &&$ 1.55    \pm 0.01    $&$ 1.3 \pm 0.3 $&&$    2.1 \pm 0.1 $&$ 3.1 \pm 1.6 $     \\
973a    &&$ 2.84    \pm 0.02    $&$ 0.75    \pm 0.08    $&&$    3.1 \pm 0.1 $&$ 0.8 \pm 0.6 $     \\
973b    &&$ 24.04   \pm 0.03    $&$.0004    \pm 0.06    $&&$    24.1    \pm 0.2 $&$ 0.1 \pm 0.7 $     \\
999 &&$ 3.939   \pm 0.003   $&$ 0.11    \pm 0.01    $&&$    4.02    \pm 0.02    $&$ 0.14    \pm 0.11$    \\
1625    &&$ 4.859   \pm 0.005   $&$ 0.15    \pm 0.02    $&&$    4.98    \pm 0.03    $&$ 0.12    \pm 0.16$    \\
1883    &&$ 1.23    \pm 0.02    $&$ 0.9 \pm 0.1 $&&$    1.59    \pm 0.09    $&$ 1.2 \pm 0.4 $     \\
2083    &&$ 8.578   \pm 0.004   $&$ 0.23    \pm 0.03    $&&$    8.67    \pm 0.03    $&$ 0.2 \pm 0.2 $     \\
2156    &&$ 14.45   \pm 0.01    $&$ 0.27    \pm 0.05    $&&$    14.64   \pm 0.05    $&$ 0.3 \pm 0.3 $     \\
3257    &&$ 2.944   \pm 0.001   $&$ 2.2 \pm 0.8 $&&$    4.5 \pm 0.3 $&$ 3   \pm 5 $   \\
3648a   &&$ 23.5    \pm 0.1 $&$ 2.7 \pm 0.6 $&&$    23.9    \pm 0.4 $&$ 1.7 \pm 4.3 $     \\
3648b   &&$ 40.99   \pm 0.01    $&$ 0.4 \pm 0.1 $&&$    41.2    \pm 0.1 $&$ 0.7 \pm 0.8 $     \\
3765    &&$66.069   \pm 0.004   $&$ 0.05    \pm 0.02    $&&$    66.12   \pm 0.02    $&$ 0.05    \pm 0.11$     \\
3891    &&$33.198   \pm 0.005   $&$-0.02    \pm 0.03    $&&$    33.24   \pm 0.02    $&$ -0.03   \pm 0.11$     \\
5478    &&$ 2.00    \pm 0.05    $&$ 0.8 \pm 0.3 $&&$    2.3 \pm 0.2 $&$ 1.3 \pm 1.7 $     \\
5563    &&$ 1.418   \pm 0.004   $&$ 0.07    \pm 0.01    $&&$    1.47    \pm 0.01    $&$ 0.08    \pm 0.06    $ \\
6581    &&$47.607   \pm 0.002   $&$0.053    \pm 0.009$&&$   47.66   \pm  0.02    $&$ 0.08    \pm 0.04$     \\
7293    &&$ 3.24    \pm 0.06    $&$ 2.0 \pm 0.9 $&&$    4.0   \pm 0.3 $&$ 3   \pm 4 $     \\

\noalign{\smallskip} \hline \noalign{\smallskip}
\end{tabular}
\end{flushleft}
\label{tab:obspeak}
\end{table*}

\subsubsection{Fluxes and Hardness Ratios}

\begin{figure}
\centering
\includegraphics[width=0.5\textwidth]{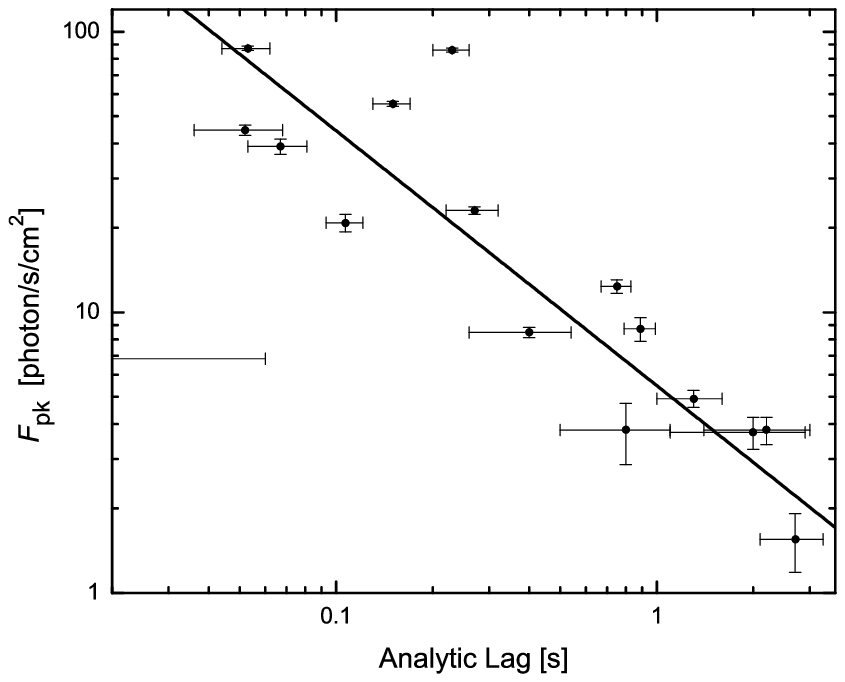}
 \caption{Peak photon flux as a function of
 spectral lag. The best fit is marked by the solid line and
approximately corresponds to an inverse relation between flux and
lag.
 \label{fig:Flag}
 }
 \end{figure}

We also calculated the  photon flux [s$^{-1}$ cm$^{-2}$] at the
peak of the light curve, $F_{\rm pk}$, which appears in column 6
in Tab. \ref{tab:HRS}.  In Fig. \ref{fig:Flag}, this flux is
plotted versus the "analytic spectral lag", $lag13$. There is one
data point that differs considerably from the power-law
correlation that appears, and that is for the second pulse in
trigger 973. Excluding this point, the best fit to the correlation
is
\begin{equation}
F_{\rm pk} = 5.44 \,  \lag^{-0.91 \pm 0.05}, \label{eq:lagflux}
\end{equation}
This relation is reminiscent of the lag-luminosity relation found
by Norris et al. (2000) and the corresponding relation for the
photon flux \citet{salm} and will be discussed further in \S
\ref{sec:FLC} . The fact that the second pulse in 973 does not
follow the general trend is of interest. As mentioned above the
lags for different pulses vary within a burst. Therefore, the
relation in Eq. (\ref{eq:lagflux}) (or correspondingly, the
lag-luminosity relation) could either be valid for all individual
pulses, each of which produces a point following the correlation,
or be  valid only for the dominant pulse in a burst. The latter
alternative was promoted by \citet{HG04} who discussed two bursts,
of which a later emission component had a significantly different
lag. Also by removing the low-intensity emission in calculating
the CCF lags, \citet{norris2} disregarded any differences in lag
for the weaker pulses. The behavior of the second pulse in trigger
973, studied here, suggests such an interpretation as well, that
is, it is mainly the dominant pulse that gives rise to the
correlation.

Finally, we calculated the hardness ratio (HR31), defined as the
ratio of the total counts in BATSE channels 3 and 1. We integrated
the count light-curves in the interval which had a flux level
above 10\% of the peak value. The ratios are given in column 7 in
Tab. \ref{tab:HRS}.

All these measurement will be compared with the high-resolution
spectroscopical analysis below (\S \ref{sec:comp}). However,
before doing this, we will investigate two issues that could
affect the interpretation, namely the use of the cross correlation
function as a measure of the spectral lag (\S \ref{sec:CCF}) and
the use of count light curves (\S \ref{sec:dc}).

  \begin{figure}[]
\centering
 \includegraphics[width=0.5\textwidth]{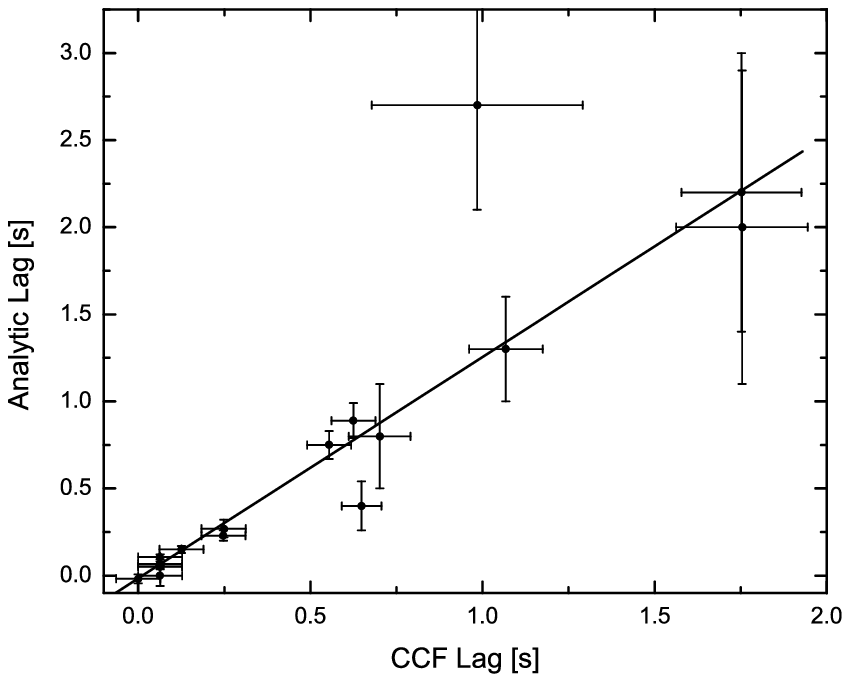}
 \caption{Analytic spectral lag, $\lag$, versus the CCF lag CCF13. A
 good linear correspondence
 exists, as expected.
 \label{fig:ccf}}
 \end{figure}

\begin{table*}
\caption[]{The HRS parameters for the studied sample}
\begin{flushleft}
\begin{tabular}{lccccccc}
\\
\hline
\\
{BATSE} & {$\eta$} & $<\alpha>$ & {$E_0$} & {$\Phi_0$}& $F_{\rm pk}$ & HR & $z^1$ \\
{Trigger} &  &  & {[keV]}  &  {[cm$^{-2}$]} & [photon/s/cm$^{2}$] & & \\
 \hline
 \vspace{-2mm}
 \\
907 &   $   2.34    \pm 0.15$&  $ 0.37 \pm0.08$ &   $504\pm75$  &   $20     \pm 1$  &       $   4.9\pm0.3   $   &     1.87    &   $ 0.40$ \\
973a    &   $   1.45    \pm 0.21$&  $ -1.01\pm0.03$ &   $515\pm25$  &   $55     \pm 11$ &   $   12.4\pm0.7   $   &     1.30  &   $ 0.35 $    \\
973b    &   $   0.84    \pm 0.1$ &  $ -1.38\pm0.06$ &   $544\pm58$  &   $17     \pm 4$  &   $   6.8\pm 0.2$   &     0.97    &   $ 1.7^3$    \\
999 &   $   1.9 \pm 0.1$ &  $ -0.5 \pm0.1 $ &   $373\pm35$  &   $24     \pm 1$  &           $   20.9\pm1.5   $   &     1.81    &   $ 0.67 $    \\
1625    &   $   2.37    \pm 0.3$ &  $ -0.74\pm0.02$ &   $1005\pm45$ &   $42     \pm 7$  &   $   55.3\pm1.1  $   &     2.35    &   $ 1.8 $ \\
1883$^2$    &   $   2.39    \pm 0.29$&  $-0.3$ -- $-2.0$    &   $451\pm21$  & $18 \pm 1$&   $   8.7\pm0.8   $   &     1.23    &   $0.45 $ \\
2083    &   $   1.78    \pm 0.04$&  $-0.36 \pm0.02$ &   $1010\pm60$ &   $65     \pm 2$  &   $   86.0\pm1.5  $   &     0.55    &   $ 0.18$ \\
2156    &   $   1.55    \pm 0.15$&  $-0.84 \pm0.02$ &   $475\pm65$  &   $42     \pm 6$  &   $   23.1\pm0.7   $   &     2.08    &   $0.41$  \\
3257    &   $   2.8 \pm 0.2$ &  $0.07  \pm 0.06$    &   $344\pm22$  &   $75     \pm 4$  &   $   3.8\pm0.4 $   &     1.65    &   $ 0.38 $    \\
3648a   &   $   0.64    \pm 0.03$&  $ -0.61\pm0.2$  &   $157\pm19$  &   $19     \pm 1$  &   $   1.5\pm0.4 $   &     0.53    &   $1.11^3$    \\
3648b   &   $   1.42    \pm 0.08$&  $ -0.36\pm0.05$ &   $716\pm17$  &   $19     \pm 2$  &   $   8.5\pm0.4 $   &     1.28    &   $ 0.38$ \\
3765    &   $   2.4 \pm 0.2$ &  $ -0.76\pm0.02$ &   $381\pm17$  &   $64     \pm 8$  &       $   45\pm2  $   &     1.86    &   $0.64 $ \\
3891    &   $   1.8 \pm 0.18$&  $ -0.75\pm0.06$ &   $269\pm13$  &   $17     \pm 1$  &       $   29.6 \pm 1.2   $   &     1.36    &   $ 0.68$ \\
5478    &   $   1.87    \pm 0.16$&  $ 0.13 \pm0.13$ &   $813\pm32$  &   $17     \pm 2$  &   $   3.8\pm0.9 $   &     1.30 &   $0.53 $ \\
5563    &   $   2.35    \pm 0.3$ &  $-0.87 \pm 0.07$    &   $320\pm15$  &   $19\pm 4$  &    $   39 \pm 2  $   &    1.35    &   $0.76 $ \\
6581    &   $   1.62    \pm 0.2$ &  $ -0.65\pm0.03$ &   $762\pm40$  &   $20     \pm 1$  &   $   87.2\pm1.5  $   &     2.04    &   $1.16 $ \\
7293    &   $   2.04    \pm 0.12$&  $ 0.79 \pm0.09$ &   $397\pm51$&   $38     \pm 4$  &     $   3.7\pm0.5 $   &   1.97    &   $ 8.6$

\\
\\

\hline
$^1$ Lag-estimated.\\
$^2$ Strong $\alpha$-evolution.\\
$^3$ Secondary pulse; ignored.
\end{tabular}
\end{flushleft}
\label{tab:HRS}
\end{table*}

\subsection{The Cross Correlation Function, $CCF13$}
\label{sec:CCF}

It is often the case, for low signal-to-noise bursts, that a
proper deconvolution cannot be made. Problems in fitting the
channel light-curves with analytical functions also arise for weak
bursts. Therefore, the cross correlation function (CCF), measured
on the count light-curves, has to be used to characterize the
shifts in time. For BATSE observations the spectral
lag\footnote{Formally the {\it lag} is the variable of the
correlation function and therefore a better notation of the
time-shift between the pulses is the {\it peak lag}.} is often
calculated between channels 1 ($\sim$ 25-50 keV) and 3 ($\sim$
100-300 keV). As the light curves (or time series) are sampled at
a number of discrete data points sums are calculated instead of
integrals (see eq. 4 in Paper I). The discrete CCF analysis was
implemented by methods similar to those employed by \citet{band97}
and \cite{norris2} and we denote this measured lag by $CCF13$. We
therefore need to investigate whether there is any difference in
the measurements between analytic lags, that is $lag13$, and
$CCF13$.

We calculated the CCF($\Delta t$, ch1, ch3) and found its maximum
value, to identify the spectral lag, $\Delta t$, between the light
curves in channels 1 and 3.   We fitted a cubic polynomial to the
peak of the resulting discrete CCF function. The statistical error
in this measurement was estimated by using a Monte Carlo routine
in which Poisson distributed noise was added to both channels
individually by an amount consistent with the observed
signal-to-noise ratio. Each iteration produced an independent lag
measurement, allowing for the accumulation of a cross-correlation
peak-distribution (CCPD). This process was continued until the
resulting CCPD approached a normal distribution, which typically
took 250-500 runs. To this resulting distribution  a Gaussian
function was fit from which the peak determines the most likely
lag and the full-width half-max yields the 1 $\sigma$ confidence
level. The results are also given in Tab. \ref{tab:lag}. In Fig.
\ref{fig:ccf} the analytic lag, $lag13$, is plotted as a function
of the CCF lag, $CCF13$. A linear fit is given which shows that
there is a good correspondence between the two methods. The best
fit is $lag13 = (-0.02 \pm 0.01) + (1.27 \pm 0.08)\times CCF13$.
The CCF lag measurements thus tend to underestimate the real time
difference between the peaks of the light curves in the two
energy bands. The bursts in our sample are simple pulse structures
and therefore the correspondence might be better than would be the
case for very variable light curves. \citet{WF00} pointed out that
it is not always reliable to determine lags with CCF. This is
especially the case for multi-peaked bursts, for which both the HR
and the CCF give average, quantitative descriptions. They also
noted that the methods used to calculate the CCFs can affect the
results considerably. For instance, the inclusion of time
intervals when the signal is at background will clearly affect the
measurements. What we find here is that the CCF measurement
represents well the actual lag for the smooth pulses in our
sample. Finally,  the $CCF13$ for the first and second pulses in
973 and the second and third pulses in 3648 are again notably
different, reconfirming the conclusion above.

\begin{table}
\caption[]{Spectral lags between channels 1 and 3 measured by
their light curves ($lag13$) and with the CCF (CCF31).}
\begin{flushleft}
\begin{tabular}{lccc}
\\
\hline
\\
 &&
\multicolumn{2}{c}{Spectral lags}
\\
\cline{3-4}

\multicolumn{1}{c}{ Trigger} && \multicolumn{1}{c}{$lag13$ [s]} &
\multicolumn{1}{c}{CCF31 [s] }
  \\

 \hline
 \vspace{-2mm}
 \\
907 &&$ 1.3 \pm 0.3    $&$1.07 \pm 0.11$\\
973a &&$ 0.75 \pm 0.08   $&$0.55  \pm0.06$\\
973b    &&$ 10^{-4}  \pm 0.06   $&$0.06   \pm 0.06$\\
999 &&$ 0.10 \pm 0.01   $&$0.06   \pm 0.06$\\
1625    &&$ 0.15 \pm 0.02   $&$0.13  \pm0.06$\\
1883    &&$ 0.89 \pm 0.10    $&$0.63  \pm0.06$\\
2083    &&$ 0.23  \pm 0.03   $&$0.25  \pm0.06$\\
2156    &&$ 0.27  \pm 0.05   $&$0.25  \pm0.06$\\
3257    &&$ 2.2 \pm 0.8    $&$1.8  \pm 0.2$\\
3648a    &&$ 2.7  \pm 0.6    $&$1.0  \pm0.3$\\
3648b    &&$ 0.4  \pm 0.1    $&$0.65   \pm0.06$\\
3765    &&$ 0.05   \pm 0.02   $&$0.06    \pm0.06$\\
3891    &&$ -0.02   \pm 0.03   $&$0.00    \pm0.06$\\
5478    &&$ 0.8 \pm 0.3    $&$0.70  \pm0.09$\\
5563    &&$ 0.07 \pm 0.01   $&$0.06   \pm 0.06$\\
6581    &&$ 0.053   \pm 0.009  $&$0.06    \pm0.06$\\
7293    &&$ 2.0  \pm 0.9    $&$1.8 \pm 0.2$

\\
\\
\hline
\end{tabular}
\end{flushleft}
\label{tab:lag}
\end{table}

\subsection{Detector Counts}
\label{sec:dc}

We now turn to the fact that the spectral lag analysis is made
straight on the count rate data, which is the method commonly used
in the literature. This introduces an important caveat since, in
general, the count light-curves have a time-dependent relation to
the photon (or energy) flux light-curves. The probability of
detecting an incoming photon is energy dependent and it therefore
becomes time dependent if the energy distribution of photons vary
with time, that is, if there is a significant spectral evolution.
For instance, for the BATSE Large Area Detector (LAD) the
probability of obtaining a count for an incoming photon is near
zero at and below 20 keV and at 30 keV it rises steeply. From 50
to 300 keV (and higher) the probability of a count approaches
unity. In the same manner, the energy measurement attached to the
count will be less correct, the higher energy the incoming photon
has; the photon is less likely to leave 100~\% of its energy in
the detector. So, to be able to interpret the count data in a
meaningful way, one therefore has to assume the spectral evolution
has negligible effect. Indeed, assuming no spectral evolution
during the pulse, there would be a constant correspondence between
the count rate in BATSE channels 2+3 and photon flux 50-300 keV
for each GRB. For a burst with a different spectrum, the constant
would be different. However, significant spectral evolution does
take place during pulses (see, e.g. Ryde (1999)) and therefore the
count-rate light-curve could result in a misleading
interpretation. Note that all these effects are well-known and are
included in the detector response matrices and are considered when
the deconvolutions are made.

We will therefore redo the analytic-lag analysis on the {\it
photon-flux} light-curves instead, but still in the same energy
bands, to see how the peak times and the lags are affected. To be
able to do this we need to deconvolve the detected count fluxes
through the detector response. The pulses in our sample are strong
enough to allow us to do this without reducing the  high
time-resolution of the $\sim 64$ ms DISCSC data. The detected
spectra were deconvolved by using the empirical {GRB model}
\citep{band}, which consists of two power laws, smoothly joined
together. The high-energy power-law, $\beta$, was fixed to the
averaged value, while the low-energy power-law, $\alpha$, was left
free to vary, whenever permitted by the quality of the data. The
purpose of the fits is to make the deconvolution and not to find
the exact parameter values and therefore we can allow quite low
signal-to-noise ratios. Finally, to arrive at the photon-flux
light curve, we integrate the photon spectra over the relevant
energy bands.

The comparison is shown in Tab. \ref{tab:obspeak}. Columns 4 and 5
contain the same information as the previous two, but now they are
measured on the photon light-curves instead. In column 4 are given
the peak times for the light-curves, integrated over the the four
energy channels, and  in column 5 are given the analytic lags. The
first conclusion to be drawn from the table is that the peak times
are systematically underestimated by using the count light-curves
compared to the deconvolved light-curves. The relative difference,
$(t_c-t_p)/t_c$, is, in general, some 10 per cent, and in some
cases much higher. Also the photon lags are, in general, somewhat
larger and they also exhibit a slightly broader dispersion.
However, even though there are noticeable differences they are not
alarming, and especially considering that larger errors are
introduced by the deconvolution process, the conclusion is that
the count light-curves, at least for the long pulses ($>$ few
seconds), do give a reasonable value for these studies.

\begin{figure*}
\centering
 \includegraphics[width=\textwidth]{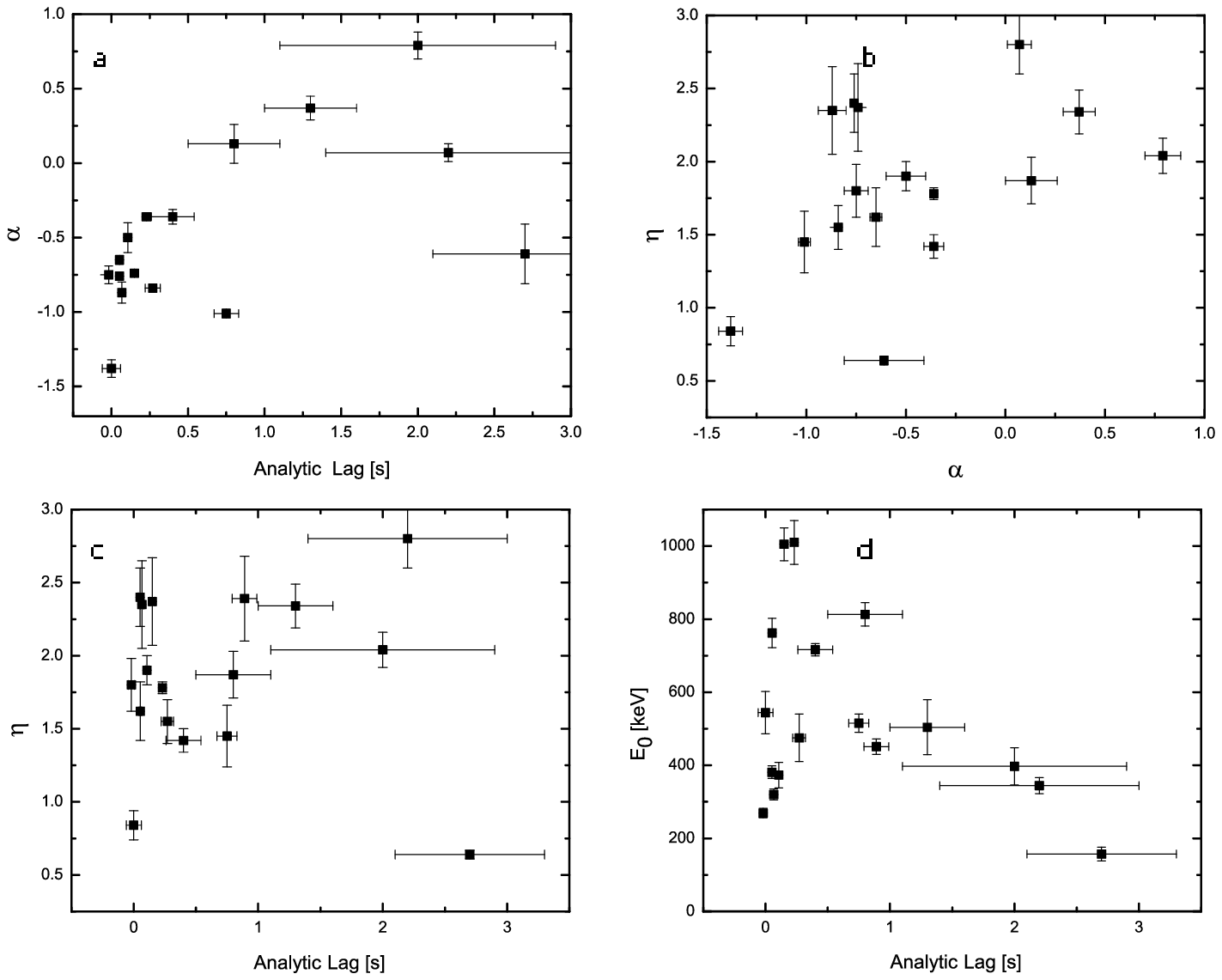}
 \caption{ {\it CGRO}-BATSE data analyzed using
 high resolution spectroscopy, HRS. a) Low-energy power-law index, $\alpha$, versus spectral lag.
 b) HIC power-law index $\eta$, versus $\alpha$. c) $\alpha$ versus lag.
 d) Initial (maximal) peak-energy,
 $\E00$, versus lag.
 \label{fig:HRS}
 }
 \end{figure*}

\subsection{Comparison with High-Resolution Spectroscopy}
\label{sec:comp}

We performed the high-spectral resolution spectroscopy on the HERB
data following the method outlined in  \cite{RS00}. We fitted the
instantaneous spectra with the highest time resolution permitted
by the data. From these fits we calculated the time-averaged,
low-energy, power-law index $<\alpha>$.  The instantaneous
$\alpha$ parameter is approximately constant in all except four
cases; triggers 999, 1625, 1883, and 2083. For instance, in
trigger 1883, $\alpha$ is constant at around -0.3 for the first
few seconds and thereafter decreases rapidly to approximately -2.
At the same time $\Epk$ decreases to $\sim 30$ keV. It is
therefore likely that the observed variation in $\alpha$ is due to
the limited energy band being used and that $\Epk$ moves beyond
the low energy threshold. The low-energy slope could therefore
very well actually be constant for this burst. From the evolution
of the instantaneous fits we analyzed the peak-energy evolution,
$\Epk=\Epk (t) $, and the spectral correlations, the
hardness-intensity correlation (HIC), $F \propto \Epk^{\eta}$ and
the hardness-fluence correlation (HFC), $d \Epk / dt = F/\Phi_0$.
The resulting parameters for the individual triggers  are given in
Tab. \ref{tab:HRS} and their dependence are shown in Figure
\ref{fig:HRS}. Panel (a) suggests that there is a trend that
harder spectra, that is larger $\alpha$, give rise to larger lags.
This is in agreement with the numerical simulations in Paper I
(their Fig. 3), in which such a trend is indeed found for all
$\eta$ values. Panel (b) shows that there is a positive
correlation, even though weak, between $\eta$ and $\alpha$; the
hard-spectral pulses have the steepest HIC power-laws. These two
relations combined suggest that $\eta$ should be positively
correlated with lag. Panel (c) shows a weak correlation in this
direction. To be able to compare this result with the numerical
simulation in Fig. 4 in Paper I, one must recognize the
correlation between $\eta$ and $\alpha$. This means that as $\eta$
increases so does $\alpha$ and the relevant panel changes from (d)
down to (a). The dominant effect turns out to be the increase in
lag due to $\alpha$, which ensures the positive correlation
between $\eta$ and lag discussed here. Finally, panel (d) shows a
trend that, at least for the large lag cases, lower $\E00$ gives
the largest lags. The numerical simulation in Paper I (their Fig.
4d) shows that this is the case for very soft-spectra pulses (for
all $\eta$). Since, for the observed bursts above, $\alpha$ and
$\eta$ are correlated [panel (b)], bursts with hard spectra
(larger $\alpha$) will predominantly have large $\eta$-values.
This means that the runs made in Paper I for large $\alpha$ (the
earlier panels in Fig. 4 in Paper I; c, b, a in increasing order)
are relevant mainly for the large $\eta$-values, i.e., the upper
section of those panels. This is indeed where the decrease in lag
as a function of $\E00$ occurs. This will cause the behavior found
in panel (d) of the figure above.

\begin{figure*}
\centering
 \includegraphics[width=\textwidth]{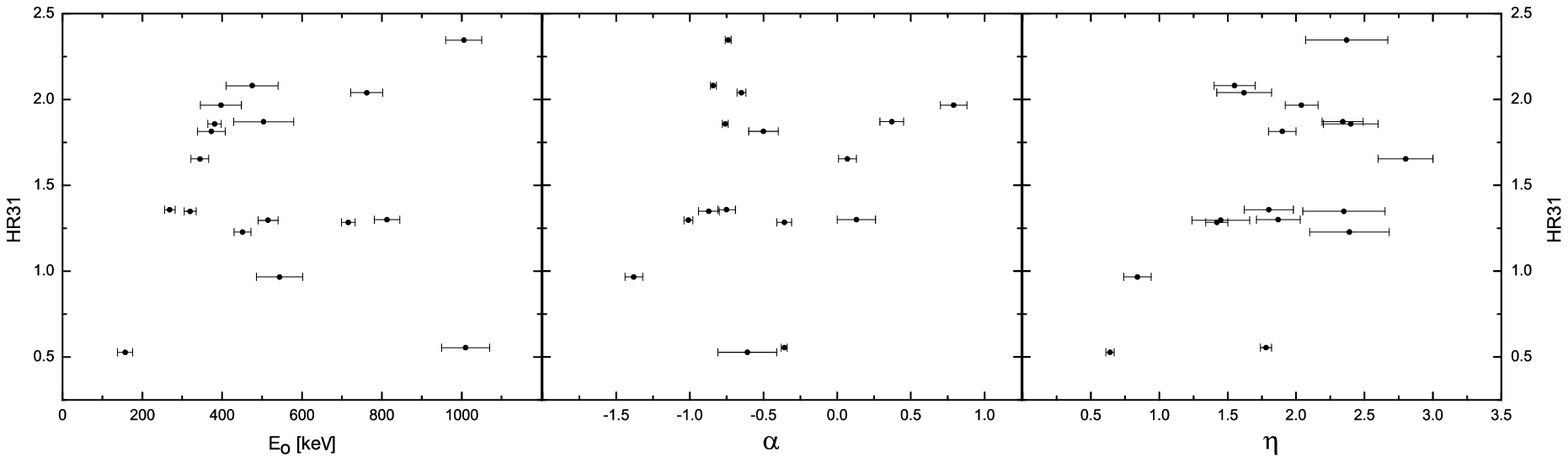}
 \caption{Hardness ratio (HR31) as a function of $\E00$, $\alpha$, and $\eta$. Only
  weak correlations emerge, as expected.
 \label{fig:HR}
 }
 \end{figure*}

In Paper I we also noted that the hardness ratios only weakly
depend on the spectral evolution parameters, since their relations
are complicated and the spectral parameters themselves have broad
dispersions which will weaken any correlations. This should be
most pronounced for low $\E00$-values (see Fig. 5 in Paper I). In
Fig. \ref{fig:HR}, we plot the measured HRs and their relationship
with the corresponding parameters $\E00$, $\alpha$ and $\eta$. The
correlations are indeed weak with only subtle trends. The panel on
the left suggests a positive correlation between $\E00$ and the
HR, which is consistent with the numerical simulations in Paper I
[their Figs. 5a and 7]. A higher initial peak energy increases the
counts in the higher channels and thereby the HRs. The two
following panels similarly suggest a positive correlation between
HR and the power-law slope, $\alpha$ and the HIC $\eta$,
respectively. According to the numerical simulations, harder
spectra (steeper HICs) are expected to have larger HR [Figs. 5
b(a) in Paper I], even though the trends are, again, expected to
be weak. A harder spectrum decreases the relative count flux in
the lower channels, thereby increasing the HRs. The HR31-$\alpha$
correlation, combined with the $\alpha$-$\eta$ correlation in Fig.
\ref{fig:HRS}d, is consistent with the HR31-$\eta$ correlation in
the panel on the right.

Finally, in Fig. {\ref{fig:KL} we plot the lag dependence on
$\Phi_0/F_{\rm tot}$. The  correlations discussed in \citet{KL03}
emerge as expected from the analytical and numerical simulations
in Paper I. The power-law index is $0.92 \pm 0.06$, consistent
with unity.

\begin{figure}
\centering
\includegraphics[width=0.5\textwidth]{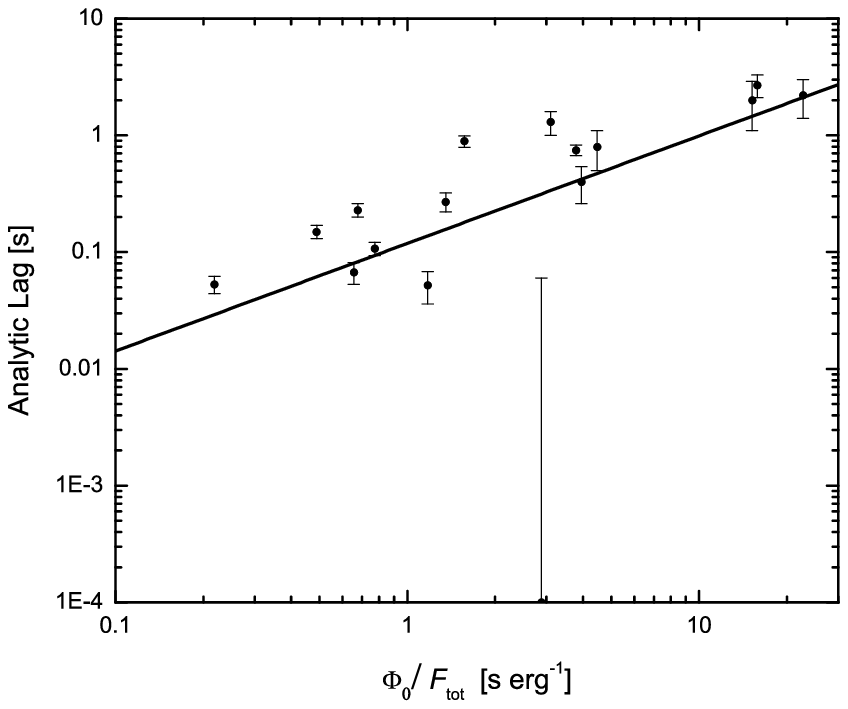}
 \caption{Dependence of spectral lag on $\Phi_0/F_{\rm tot}$. The outlier point is from the second pulse
in trigger 973.
 \label{fig:KL}
 }
 \end{figure}

\section{Discussion}
\label{sec:physics}

Interestingly, equation (\ref{eq:lagflux}) shows  that the
observed photon (energy) flux of a pulse is inversely proportional
to the spectral lag. \citet{norris2} similarly found that bursts
with high peak-fluxes predominantly have small lags (their Fig.
2). However, the flux lag relation we find is a much tighter
correlation than compared to the plots in \citet{norris2}. Our
analyses differ in several aspects. \cite{norris2} study a general
sample (including bursts with lower peak fluxes) and not only
single pulses. The lag value for multi-pulsed events will be an
averaged value and the HR rather catches the distribution of
$\E00$-values among the pulses in the studied burst. The general
distribution in the flux-energy plane is furthermore determined by
the HIC parameter $\eta$.  Using bursts with a known redshift,
\cite{norris2} translated their fluxes to luminosities and a
stronger correlation emerged. More specifically, they found that
the spectral lag, $\Delta t$, and the {\it isotropically}
equivalent peak luminosity [or luminosity per steradian, that is,
the power output of the burst assuming that the explosion occurs
isotropically], $\Liso$, [erg/s] (see also \citet{ SDB, salm})
follows the relation
\begin{equation}
\Liso = 2.9\times 10^{51}{\mathrm{ erg/s }}\left(\frac{\Delta
t}{0.1 {\rm{ s}}}\right)^{-1.14\pm0.20}.  \label{eq:LT}
\end{equation}
Using the photon-number peak-luminosity [photons/s] instead,
\citet{salm} found the exponent to be $-0.98$, which all together
is suggestive of inverse linear proportionality. We will in \S
\ref{sec:FLC} below argue that Eq. (\ref{eq:lagflux}) most
probably is a consequence of this relation.

A positive correlation between the peak energy of the spectrum,
$\Epk$, and the luminosity has been suggested empirically
\citep{LRR02, amati}. Combining this with the lag-luminosity
anti-correlation (Eq. \ref{eq:LT}) an anti-correlation is inferred
between peak-energy and lag. This is indeed consistent with the
behavior of the pulses in our sample, shown in Fig.
\ref{fig:HRS}d, especially for larger lags. The bursts with the
longest lag have the lowest peak energies. Furthermore, the
positive correlations between $\alpha$ and $\eta$ with lags (Figs.
\ref{fig:HRS}a and c) similarly imply that harder spectra (large
$\alpha$) as well as steeper HICs (large $\eta$) are associated
with bursts of lower luminosity.

We also noted that different pulses within a burst have different
lags. Such a conclusion was similarly drawn by Hakkila et al.
(2004). Finally, we also note that \citet{norris2} found that
bursts with large HRs predominantly have small lags. For our
pulses in Tab. \ref{tab:obspeak} we similarly find that the
largest HR bursts mainly have small lags.

  \begin{figure}[]
\centering
 \includegraphics[width=0.5\textwidth]{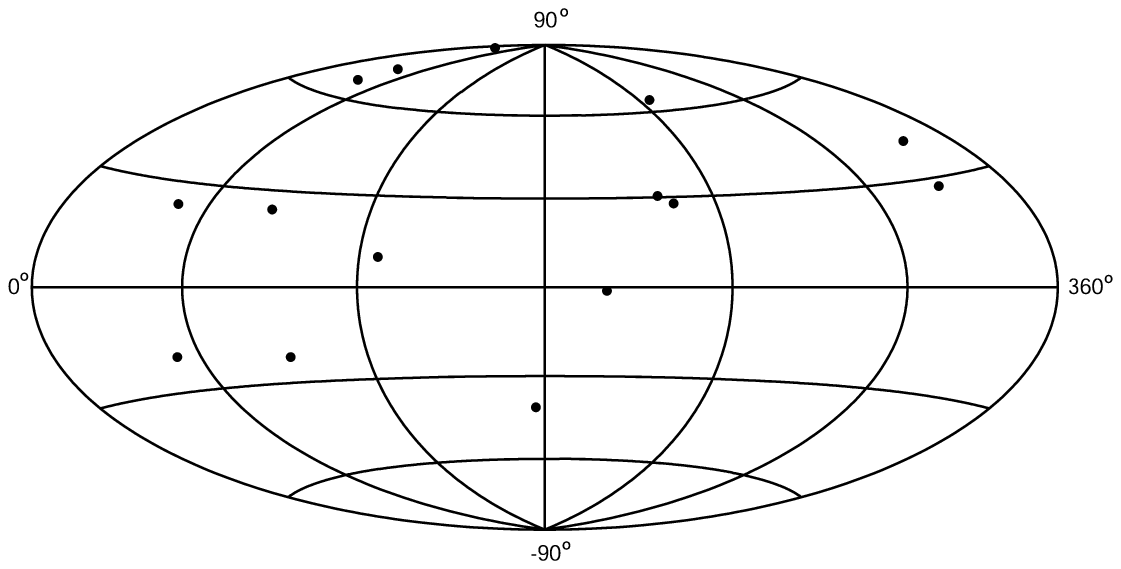}
 \caption{Sky distribution in super-galactic coordinates for the
 15 BATSE bursts in the studied sample.
These are characterized by having long uncontaminated FRED
pulse(s). The dipole moment of the distribution is very large.
 \label{fig:sky}}
 \end{figure}

\subsection{Origin of the Flux-Lag Correlation}
\label{sec:FLC}

We find, in our sample, that the observed flux dispersion is
$\sim$ 1.5 orders of magnitude and follows a lag-flux power-law
relation (Eq. [2]). Comparing it with the similar relation between
the luminosity and lag of \citet{norris2} we can, in principle,
have two different possibilities. First, the lag-flux relation is
a consequence of the lag-luminosity correlation. In this case, the
latter relation has to be valid for our sample as well. Second,
Eq. (2) could be dominantly a consequence of the distribution of
the 15 bursts in redshift. In this case, Eq. [4] may, but need
not, be correct. Since the dispersion in the measured fluxes is
1.5 orders of magnitude, it is obvious that for the former case a
smaller dispersion in $z$ is needed. Below, we will argue that it
is indeed the lag-luminosity relation that is revealed.

The quantity measured by BATSE is the photon count rates and the
derivation of the energy fluxes will by necessity be model
dependent. Therefore it is more useful to use the more fundamental
quantities given by the photon rates in discussing these matters.
The photon luminosity, $L_{\rm iso}$ [photon/s], that corresponds
to a certain photon flux value, $F$ [photon/s/cm$^{2}$], observed
from a source at a redshift, $z$, is given by \citep{CPT92} (see
also \citet{MM96})
\begin{eqnarray}
L_{\rm iso} & =& \frac{4 \pi F}{1+z} \,\, d_{\rm L}^2\, ;\label{eq:z2}\\
 d_{\rm L} &= &
\frac{ (1+z) c}{H_0} \int^z_0 \frac{dz'}{\sqrt{(1+z')^2
(1+\Omega_M z') - z'(2 + z')\Omega_{\Lambda}}} \label{eq:z}
\end{eqnarray}
where $H_0$ is the Hubble constant, $\Omega_M$ is associated with
the present day matter density, $\Omega_{\Lambda}$ with the dark
energy density, and $\Omega_M + \Omega_{\Lambda} = 1$, that is, we
consider a flat universe at once. The luminosity distance is
denoted by $d_{\rm L}$. The extra $(1+z)$ factor in the
denominator in Eq.(\ref{eq:z2}) account for the fact that we are
using the photon flux and not the energy flux which is assumed in
the definition of $d_{\rm L}$ (see, e.g, end of section 2 in
\citet{M&M95}). Furthermore, in Eq. (\ref{eq:z2}) we have
neglected the $K$-correction, which takes into account the effect
of $z$ on the bandpass. As shown by, for instance, \citet{bloom},
the quantitative distribution of fluxes does not change
dramatically.

For the flux-lag correlation to emerge, the dispersion of the
quantity $(1+z)/d_L^2$ should then be smaller than $\sim$ two
orders of magnitude. If this is the case, then the dispersion in
luminosities should dominate the dispersion of flux. It has
actually already observationally been confirmed that such
dispersions exist. For the luminosities of bursts with known
redshifts there is a dispersion of roughly two-to-three orders of
magnitude (e.g., Fig. 6 of \citet{M01}). Taking typical values of
$\Omega_M = 0.27$ and $\Omega_{\Lambda} = 0.73$  \citep{riess04}),
and if the pulses originate in bursts that are, say, from $0.5 < z
< 1.0$, then the ratio of minimal and maximal $d_L$ is $5.5$, and
hence $(1+z)/d_L^2$ varies by a factor $\simeq 20$. Note that for
this ratio the exact value of $H_0$ is unimportant. With such a
distribution in $z$, the lag-luminosity relation of Eq. [4] will
be dominant, and the lag-flux correlation will be observed. The
actual distribution in $z$ is not known, but more arguments can be
found to support such a range.

First of all, it must be recognized that, remarkably, the quantity
$(1+z)/d_L^2$ has the largest dispersion for the smallest
$z$-values \citep{MM88}. If our analyzed bursts were at say $z
\simeq 0.005$ one would thus expect a large dispersion diluting
the lag-luminosity correlation. Furthermore, in the case of these
small redshifts one would also expect a concentration of objects
towards the galaxies in the local universe. These are distributed
in a major planar structure centered around the Virgo cluster of
galaxies, called the super-galactic plane, SGP \citep{Peebles,
SuperG, norris3}. If the bursts in our sample were to be
associated with these galaxies, their distribution in
super-galactic coordinates would have a high quadropole moment,
and the conclusion would be that they lie within a redshift of $z
\sim 0.01$. We note here that only one GRB has been detected at
 a similarly small distance (GRB 980425/SN98bw, $z = 0.0085$). We
therefore plot in Fig. \ref{fig:sky} the sky distribution of our
sample in super-galactic coordinates. We calculated the dipole, $D
= <\cos(sgb)>$, and quadropole, $Q = <\sin^2(sgb) -1/3>$ moments
of the angular distributions. For the 15 bursts, $D = 0.362$
(notably large) and $Q = 0.0455$. These bursts are avoiding the
south super-galactic region, hence the large dipole moment. Since
the belt around the super-galactical equator, up to $sgb = \pm 30
\deg$  in super-galactical latitudes, covers half the sky, one
would expect a statistically significant over-density in this belt
for objects associated with the SGP. However, the quadropole
moment we find is relatively low; in the $sgb = \pm 30 \deg$--belt
there are 8 GRBs, in the remaining part of sky 7 GRBs. In the
BATSE exposure function, in the super-galactic plane, there will
be higher, non-vanishing moments, albeit they will be small, and
at least for 15 events the shot noise will cover them. We also
produced 1000 Monte-Carlo catalogues with 15 events to test the
robustness of this result. These catalogues follow the BATSE sky
exposure. If we order the simulated distributions from 1 to 1000
in increasingly anisotropic distributions, the observed dipole
value lies at 993, that is, the confidence limit is about 99.3\%.
Our sample is thus not strongly distributed along the
super-galactic plane and we therefore conclude that our sample is
not completely within this distance, but beyond it. In this
connection it should be noted that \cite{norris3} studied a
(totally different) sample of long-lagged burst and plotted them
in super-galactic coordinates and did indeed find a large
quadropole moment. This led him to suggest that his sample is
associated with SNe type Ib/c in the local super cluster of
galaxies.

Second, very large $z$-values for the 15 GRBs in our sample can
also be excluded. Our sample is limited to the brightest BATSE
bursts ($> 2$ photons/s/cm$^2$) and to cases with a single or a
few well shaped pulses. The first constrain leaves bursts from the
BATSE catalogue which are normally totally dominated by lags which
are very short (${\ltsima} 250$ ms; see Fig. 2 in \cite{norris3}).
According to Eq. (\ref{eq:LT}) these short-lag bursts correspond
to luminous explosions leading to the high fluxes observed. The
\citet{norris3} sample also has the property that bursts with lags
greater than $\sim$3 s all have peak fluxes lower that 2
photons/s/cm$^2$. However, the measured lags in our sample (Tab.
\ref{tab:lag}) are dominated by long lags and are thus
low-luminosity bursts (according to Eq. [\ref{eq:LT}]). Following
the nomenclature introduced by \cite{norris3} these thus
constitute a fourth region in the peak-flux--lag plane, namely the
bright and long-lag bursts. Since they are low-luminous bursts but
still bright this means that they cannot be very distant. The
second constraint that we used excludes the most variable bursts.
According to the empirical correlation between variability and
luminosity (\cite{reich}) this again leaves low-luminosity bursts.
To be able to study individual pulses we need bursts of low
variability which are also bright. These criteria left us with
this sample.

Third, by assuming the lag-luminosity relation to be valid for our
sample, we can calculate a luminosity for each pulse and, in
combination with the measured flux value, we can arrive at an
estimation of the redshift. These {\it estimated} redshifts  are
given in the last column in Tab. \ref{tab:HRS}. The redshifts are
all below $z \sim 1.8$ except trigger 7293, which is derived to
have $z = 8.6$. The peak of the distribution is at $z \sim 0.7$.
In any case, this also supports the small dispersion of
$(1+z)/d_L^2$. Note here the obvious discrepancy in redshifts for
the multi-peaked events which arises due to the variation in lags,
already pointed out in papers I and above. The sample studied is
therefore most probably from a relatively limited distance range
(low dispersion in $z$) at not very high redshifts, but mainly
beyond $z \sim 0.01$.

Another argument that supports the contention that the
correlations in Eqs. (2) and (4) are the same comes from the
cosmological time-dilation. Let us rewrite Eq. (4) in the form
$$L/\Omega = L' (t_0/\Delta t_{\rm rf})^{\alpha},$$
where $L' = 2.9\times 10^{51}$ erg/s, $t_0 = 0.1$ s and $\alpha
\simeq 1$. For our purpose it is essential that $\alpha > 0$, and
the fact that $\Delta t_{\rm rf}$ is the rest-frame lag. Then the
observed flux will be given by $F = L' (t_0/\Delta t_{\rm
rf})^{\alpha} (1+z)/d_L^2 = L' (t_0/\Delta t_{\rm obs})^{\alpha}
(1+z)^{1+\alpha}/d_L^2 $, since $\Delta t_{\rm obs} = (1+z) \Delta
t_{\rm rf}$. This means that one should consider the dispersion of
$(1+z)^{1+\alpha}/d_L^2$ instead of $(1+z)/d_L^2$. A positive
$\alpha$ value will decrease the effect of the distance dispersion
on the correlation.

In summary, all this suggests that the correlation between the
observed fluxes and lags (Eq. [2]) is a residual of the
lag-luminosity correlation (Eq. [\ref{eq:LT}]) that is still
apparent.

\subsection{Origin of the Spectral Correlations}

Several observational facts appear to indicate that long GRBs
originate when the iron-core of a massive star collapses to a
black hole or a rapidly rotating neutron star, producing a
collimated, relativistic jet (collapsar model; see \citet{W93}).
The collimation angle of the outflow varies, which leads to
different behaviors of the observed light curves. The observer is
assumed to see the GRB approximately along the jet axis and the
opening angle of the collimation of the outflow, $\Tj$, may vary
and is responsible for the diversity in the quantity $\Liso$.
Since $\Liso$ is observed to be correlated to the variability of
the $\gamma$-ray light curve, $V$ \citep{reich}, a
$V-\Tj$-correlation should also exist. This was indeed shown to
exist by Kobayashi, Ryde, \& MacFadyen (2002), who used bursts for
which an opening angle had been deduced. They also suggested an
explanation of the correlations within the internal shock model.
There are two possible smoothing effects and they both dependent
on the bulk Lorentz factor: the location of the pair photosphere
produced by synchrotron photons at $R_{\pm}\sim 1/\sqrt{\Gamma}$
and the angular time-scale $t_{\rm ang} \sim R/2c\Gamma^2$. If a
collision happens below the photosphere, the whole internal energy
produced by the collision is converted to kinetic energy again via
the shell spreading. It is only collisions that occur above the
photosphere that can be seen by an observer. For instance, a small
$\Gamma$ yields a large photospheric radius and the outflow then
becomes more ordered and therefore the variability is lower. In
the same manner, a small $\Gamma$ yields a large $t_{\rm ang}$ and
thus a less variable light curve. Therefore, the authors argued
that the variability is determined mainly by the bulk $\Gamma$ of
the outflow and thus the $V-\Tj$ correlation reduces to a $\Gamma
- \Tj$ correlation. Simulating the hydrodynamic collisions
numerically, \citet{KRM} found that the data are fitted the best
with $\Gamma \propto \Tj ^{-1}$ and thus the mass loading $M
\propto E/(c^2 \Gamma) \propto \theta$, since the total explosion
energy has been found to be approximately the same from burst to
burst (e.g., \citet{fra01}). This scenario is reasonable in the
collapsar model: a wide jet leads to a dim burst since a wider jet
involves more mass at the explosion, leading to lower speeds of
the outflowing material. The change in $\Liso$ therefore depends
mainly on $\Tj$. The variability, $V$, is determined by the
opening angle which in its turn also affects the quantity, $\Liso
\propto L_{const}/\Tj ^2 $. In this model $L \propto \Gamma^2$.

If the spectral lag is also determined by the angular spreading
time scale, which is the key time scale in this  model, the
lag-luminosity relation can be found. In this model luminosity is
scaled by the opening angle $L \propto \theta^{-2}$, while the
angular spreading time is $t_{ang}\propto \Gamma^{-2} \propto
\theta^2$. This leads to  $L \propto t_{ang}^{-1}$, which is Eq.
(\ref{eq:LT}). According to Paper I, the lag is mainly determined
by the pulse time scale $T$, which is defined as $T = \P0
\E00/F_0$. Since lag and luminosity are correlated, $\P0$/flux
should also be approximately correlated, as shown in Fig.
\ref{fig:KL}.

In the description above it is assumed that the variations in
Lorentz factors are mainly due to  variations in the collimation
opening-angle of the outflow, and that the jets have a uniform
profile out to the opening angle. However, an alternative
scenario, for which the discussion above is also applicable, is
the structured jet in which a certain beam profile is assumed
(e.g. \citet{rossi}).  In such a case the most luminous part of
the jet has a higher Lorentz factor. A large jet opening angle
gives the possibility for large off-axis views. The probability
for a certain viewing angle is indeed proportional to the viewing
angle itself, and thus large angles are favored. If there is a
typical jet profile in all bursts, larger opening angle  outflows
are most often seen more off-axis, with a lower $\Gamma$ as a
consequence.

\citet{ryde04} found, by analyzing spectrally hard BATSE pulses,
that the spectra can be modelled by a thermal black-body
superimposed on a non-thermal, optically-thin synchrotron spectrum
(see also Ghirlanda et al. 2003). The relative strengths of these
two components vary. A large $\alpha$-value is measured if the
thermal component is dominant, while a lower value, more like a
synchrotron value, is found if it is less prominent and the
non-thermal component dominates. Small outflow velocities,
$\Gamma$, lead to larger photospheric radii, and less dissipation
of energy in shocks (for example) which enhances the thermal
emission, thus leading to a harder $\alpha$, as well as larger
lags. So, a low $\Gamma$ leads to low luminosity bursts (due to
larger photospheric radii, as well as lower $\Gamma$) and hard
spectra. This is indeed the trend in Fig. \ref{fig:HRS}a, assuming
that the lag can be translated to luminosity through the
lag-luminosity relation. Here, that relation is observationally
based on Fig. 2 and all the relations in Fig. \ref{fig:HRS} then
correspond to relations with luminosity. A pure black-body
emission should have a HIC of $\eta = 4$. If the photospheric
emission is dominant, a high $\eta$ should be correlated with a
lower luminosity (Fig. \ref{fig:HRS}c). This also provides a
natural explanation for a $\alpha - \eta$ correlation in Fig.
\ref{fig:HRS}b.

Further, Fig. \ref{fig:HRS}d suggests that for low-luminosity
bursts there is a positive luminosity-to-peak-energy correlation.
For the smallest lags (highest luminosities) the peak-energies
decrease and there is an anti-correlation.  In the optically-thin
synchrotron-shock model an anti-correlation arises most naturally.
$\Epk \propto \Gamma B' \gamma_e^2$, where $B'$ is the comoving
magnetic-field strength and $\gamma_e$ is the characteristic
Lorentz factor of an emitting electron, and since, in the internal
shock model, $\gamma_e^2$ is mainly dependent on the relative
Lorentz factor of the colliding shells and not on $\Gamma$, $\Epk$
is proportional to $\Gamma B'$. This is the Lorentz-boosted
magnetic field strength, of which the square  should be
proportional to the fireball energy density in the observer frame,
$U \propto L/R^2$, where $R \propto \Gamma^2$ is the typical
collision radius. Here we have assumed that the magnetic energy
density is a constant fraction of the total energy density.
Therefore we have $\Epk \propto L^{1/2} R^{-1} \propto L^{1/2}
\Gamma^{-2}$. With, for instance, $L \propto \Gamma^2$ from above,
we have $\Epk \propto L^{-1/2}$, that is, an anti-correlation.
This is for synchrotron radiation. The peak energy of
inverse-Compton scattered photons is at $\Epk^{\rm IC} =
\gamma_e^2 \Epk^{\rm synch}$ and thus it has the same dependence.
\citet{ZM02} also discussed the luminosity--peak-energy
correlation for various other outflow models: For  Poynting-flux
dominated models $\Epk \propto \Gamma$, while for emission from
the baryon photosphere, during the shell acceleration phase, $\Epk
\propto L ^{1/4}$. The last two outflows thus easily reproduce
positive correlations. Furthermore, \citet{ryde04} motivated that
the time-evolution of the temperature decay of the thermal pulses
they studied can be explained either by a baryonic photosphere
during the acceleration phase or by a Poynting-flux dominated
wind. In both these models a positive correlation is, as
mentioned, expected. We can therefore interpret Fig.
\ref{fig:HRS}, following the suggestions in \citet{ryde04}. The
positive correlation for low luminosities is for pulses that are
dominated by the thermal emission. The negative correlation (at
small lags) are then from pulses in which the non-thermal,
synchrotron component dominates and thermal emission is less
dominant. This is indeed consistent with Fig.\ref{fig:HRS}a in
which the small lag pulses have soft spectra, while the large lag
cases have harder spectra.

This simple description can thus accommodate the main correlations
and trends presented above. A burst does not necessarily need to
have the same lag for all pulses within it, since it is mainly
internal properties that determine the characteristics.

\subsection{Alternative Models}

 Throughout this paper it has been emphasized that the $\Epk$
evolution has an important role in determining the spectral lags.
The HFC, $\dot{\Epk} = -{F}/{\Phi_0}$, which governs this
evolution, could be due to thermal processes like saturated
Comptonization without extra heating, which is difficult to
achieve in general however. Such a scenario has been discussed by
Liang \& Kargatis (1996) and Liang (1997). \cite{schaefer04}
argued further that the lag-luminosity relation can be explained
by this behavior. This model assumes an impulsive heating of a
confined plasma which starts to cool. There should not  be any
continuous heating, for example, dissipation of energy through
shocks. Furthermore, there has to be a constant injection rate of
soft photons which gives rise to the cooling of the electrons,
$\dot{T_e}$. The thermal emission ensures that the emitted photons
have approximately the same energy as the averaged electron. The
change of internal energy of the cloud $\dot{U}$, which gives the
luminosity, is then $L = \dot{U} = N \dot{T} = N \dot{ \Epk }$,
where $N$ is the number of electrons. This equation is consistent
with the HFC. Assuming, first, that $d\Epk/dt$ is constant for a
pulse (which is not the exact case, see \cite{RS02}), second, that
the peak energy decays within typical range, that does not differ
greatly from burst to burst, (clearly also an approximation), and,
third, that the peak luminosity, $L_{\rm peak}$ is a good measure
of the pulse luminosity, $L$, then the analytical finding in Paper
I that
\begin{equation}
\Dt = \int_{\Delta E} \frac{dt}{d\Epk} \,dE
\end{equation}
combined with the HFC, gives that the lag
 \begin{equation}
 \lag \propto - \frac{\P0}{L_{\rm peak}}.
 \label{eq:31}
 \end{equation}
$\P0$ is practically constant between pulses, and in the model
above it is a combination of $N$, distance,  and jet opening
angle. This is similar to  the lag-$\P0$ correlation, discussed by
\cite{KL03}. This model has several advantages but needs a number
of approximations and strong assumptions to hold.

As mentioned above the lag-luminosity relation is intriguingly
close to a pure proportionality between the luminosity and the
inverse of the lag. \cite{salm} noted that such a relation follows
from simple relativistic kinematics if one assumes a common
comoving time, which could be due to cooling or deceleration.
Assuming first that the viewing angle (between the jet velocity
and the line-of sight, l.o.s.) $\theta$ is not large, the
angle-dependent Lorentz factor ${\cal{D}(\theta)}$ can be
approximated by $2 \Gamma$ and the photon-number luminosity goes
as $F_N = \Gamma F_N' (1 + z)^{-1}$, where $F_N'$ is the comoving
value. Since a typical comoving time scale, $\delta t$ goes as
$\delta t' (1+z)/2\Gamma$, the observed lag-luminosity relation
emerges. The dominant effect in the dispersion of lags is assumed
to be due to the value of $\Gamma$ directed towards the observer.
This could be explained in two different ways. First, it is
natural to assume that the Lorentz factor is different from burst
to burst. Second, if the outflow is in the form of a jet, the flow
velocity could very well be dependent on the angle from its axis,
as mentioned above. If the l.o.s. is along the jet axis a large
$\Gamma$ is observed, while if the l.o.s. is different from the
jet axis the flow velocity might be significantly lower. The
transformation of the time-scale and the flux will depend on the
viewing angle, since they are affected by the angle-dependent
Lorentz factor, ${\cal{D}} (\theta)$. \cite{nakar} even argued
that it is indeed the viewing angle that causes the dispersion in
spectral lags and they showed that they could reproduce the
observed relation between the peak luminosity and lags.

\cite{DM03} presented a pulse description within the internal
shock model, which can reproduce the lag-luminosity relation. The
peak energy is parameterized as $\Epk \propto \rho^x \epsilon^y
\Gamma$, where $\rho$ is the post-shock density and  $\epsilon
c^2$ is the post-shock dissipated energy per unit mass. Standard
synchrotron radiation, assuming classical equipartition, has
$x=1/2$ and $y=5/2$. However, if the equipartition between the
magnetic field and particle energies vary with either $\rho$,
$\epsilon$, or both, then other values of $x$ and $y$ can exist.
The authors found that the observed spectral pulse behavior is
best reproduced with $x = y = 1/4$.  Their model also predicts
that the hardness-ratio should decrease with spectral lag, which
is similar to what is found by \cite{norris2}.

\section{Conclusions}
\label{sec:discussion}

We have studied the spectral lags in a sample of bright {\it
pulses}, all belonging to the class of long duration bursts (see,
e.g. \citet{B03, H98}). We find that there is a large spread in
the measured lag values. In particular we find that the lag values
differ considerably between pulses within a burst. This is
important to note, since often the whole light curve is used to
measure the spectral lag, and the lag measurement then by
necessity is only an averaged value. Also, models which cannot
accommodate a variable lag over a pulse, that is, models which
rely solely on external effects, need to be refined.

We also find that a correlation between the observed flux and the
spectral lag is valid for prominent pulses, very similar to the
lag-luminosity relation that has previously been reported for {\it
entire} bursts, that is, not only for pulses. We draw the
conclusion that the observed relation is a residual of the
physical lag-luminosity relation. The sample does not show a
significantly strong clustering towards the super-galactic plane,
which indicates that the bursts are at a larger distance than $z
\sim 0.01$. However, they cannot lie at very large distances,
since they are mainly low-luminous and bright bursts.

We have compared the analytic lags with the values found by using
the CCF. The CCF underestimates the lags somewhat, but there is a
linear relation between the measurements. Also we find that the
results found by using the count light-curves and the photon-flux
light-curves do not differ significantly and thus permit their
use.

The observed correlations are similar to the general analytical
and numerical simulation results in Paper I. Foremost we find that
the lag correlates strongly with the decay time-scale of the pulse
and that the relation is linear. This means that the relationships
including the lag, that have been identified, should be
translatable to relationships including the pulse time-scale. This
time-scale is closely connected to the creation process and
radiation processes in the out-flowing plasma.

We also argue for a collapsar model scenario, where the
collimation is the important factor leading to variation in the
observed parameters. The observations are consistent with the
existence of several components in the spectra, thermal and
non-thermal emission, varying in relative strength.

\begin{acknowledgements}
We are grateful to Drs. M. Briggs and R. Preece for valuable
support. We also thank Drs. C.-I. Bj\"ornsson, S. Kobayashi, and
P. M\'esz\'aros for interesting discussions. The anonymous referee
is acknowledged for useful comments that improved the manuscript.
This study was supported by the Swedish Research Council and the
Hungarian OTKA grant No. T034549, by the research plan J13/98:
113200004 of the Czech Ministry of Education, Youth and Sports,
and by a grant from the Wenner-Gren Foundations (A.M.). This
research made use of data obtained through the HEASARC Online
Service provided by NASA's Goddard Space Flight Center. FR wishes
to express his gratitude to the host Departments in Prague and
Budapest for their kind hospitality during his visits.
\end{acknowledgements}

\end{document}